\pdfoutput=1 
\documentclass[a4paper]{article}
\usepackage{INTERSPEECH2021,amsmath,graphicx,tipa,url,tikz,tabularx}
 \usepackage[T1,T2A]{fontenc}
\usepackage[utf8x]{inputenc}
\usepackage[ukrainian,english]{babel}
\usepackage{etoolbox}
\usetikzlibrary{positioning,shapes.multipart,calc,arrows.meta}
\usepackage{placeins}
\usepackage{tempora}
\usepackage{tipa}
\usepackage{cite}

\title{SoundChoice: Grapheme-to-Phoneme Models with Semantic Disambiguation}
\name{Artem Ploujnikov$^{1,2}$, Mirco Ravanelli$^{1,2,3}$}
\address{
  $^1$Mila - Quebec AI Institute \\
  $^2$ Université de Montréal\\
  $^3$ Concordia University}
\email{artem.ploujnikov@umontreal.ca, ravanellim@mila.quebec}

\begin{document}

\maketitle
\begin{abstract}
End-to-end speech synthesis models directly convert the input characters into an audio representation (e.g., spectrograms). Despite their impressive performance, such models have difficulty disambiguating the pronunciations of identically spelled words.
To mitigate this issue, a separate Grapheme-to-Phoneme (G2P) model can be employed to convert the characters into phonemes before synthesizing the audio. 

This paper proposes \textit{SoundChoice}, a novel  G2P architecture that processes entire sentences rather than operating at the word level. The proposed architecture takes advantage of a weighted homograph loss (that improves disambiguation), exploits curriculum learning (that gradually switches from word-level to sentence-level G2P), and integrates word embeddings from BERT (for further performance improvement). Moreover, the model inherits the best practices in speech recognition, including multi-task learning with Connectionist Temporal Classification (CTC) and beam search with an embedded language model. As a result,  SoundChoice achieves a Phoneme Error Rate (PER) of 2.65\% on whole-sentence transcription using data from LibriSpeech and Wikipedia.
\end{abstract}


\section{Introduction}

Speech synthesis systems convert written text into a sequence of speech sounds. The irregularities commonly encountered in natural language orthography pose significant challenges to this process. For instance, a given sequence of characters (grapheme) can yield different pronunciations depending on the context (homographs).  The sentence "English is t\textbf{ough} [\textipa{2f}]" can be understood thr\textbf{ough} [[\textipa{u:}] thor\textbf{ough} [\textipa{@}] th\textbf{ough}t [\textipa{A}] th\textbf{ough} [\textipa{oU}]. In some cases, the disambiguation depends on parts of speech (live - [\textipa{laIv}] vs [\textipa{lIv}]) or semantics (bass - [\textipa{beIs}] vs [\textipa{b\ae s}]).
Popular end-to-end speech synthesis models often fail to perform disambiguation of the homographs. Tacotron \cite{tacotron}, for instance, is successful at only the most basic disambiguation (e.g. "read" - past vs present), while DeepVoice3 \cite{ping2018deep} produces intermediate phonemes in homographs.

Grapheme-to-Phoneme (G2P) models can improve the system's performance in these cases. Several approaches have been proposed in the literature: early attempts were mainly based on classical methods (e.g., Hidden Markov Models\cite{g2p-hmm-taylor}), while 
more modern approaches rely on sequence-to-sequence deep learning.
LSTM-based models \cite{g2p-lstm-rao} have been largely adopted for this task,  and, more recently, transformer-based models \cite{transformer-g2p} and convolutional models \cite{conv-g2p} have been proposed as well. 
These models are typically trained and evaluated on word-level lexicons (e.g.,  \text{CMUDict}\cite{cmudict}), making it impossible to resolve homograph disambiguation. 
The task of homographs disambiguation, on the other hand, has been explored in the literature as an independent research direction.
Indeed, it was mainly framed as a classification task rather than an actual Grapheme-to-Phoneme conversion. 
Early work includes a classical hybrid method combining a rule-based algorithm and multinomial classifiers, such as the method proposed by Gornman et al.,  \cite{gorman-etal-2018-improving}. A BERT-derived classifier model based on contextual word embeddings\cite{homograph-contextual-word-embeddings} has been proposed as well. A recent example of a model exploiting sentence context is T5G2P\cite{t5g2p}. DomainNet \cite{domainnet}, instead, handles the task from a purely semantic view,  while Alqahtani et al. \cite{homograph-contextual-word-embeddings}  propose a self-supervised method for languages with diacritics that are frequently omitted.

This paper introduces \textit{SoundChoice}, a novel G2P model that builds on insights from earlier contributions and addresses some of their prominent limitations. Different from previous methods, SoundChoice operates at the sentence level. This feature enables the model to exploit the context and better resolve homograph disambiguation. To further improve disambiguation, we propose a homograph loss that penalizes errors made on homograph words. The homograph disambiguation is not framed as a separate classification problem but is embedded into the G2P model itself through our homograph loss. In summary, the proposed SoundChoice introduces the following new features:

\begin{itemize}
    \item It works at a sentence level, and it is trained with a weighted homograph loss.
    \item It gradually switches from word- to sentence-level G2P using a curriculum learning strategy.
    \item It models the sentence context by taking advantage of a mixed representation composed of characters and BERT word embeddings.
    \item It introduces Connectionist Temporal Classification (CTC) loss on top of the encoder and combines it with the standard sequence-to-sequence loss computed after the decoder (as commonly done in speech recognition).
\end{itemize}

Our best model achieves competitive Phonene-Error-Rate (PER\%) on LibriSpeech sentence data (best test PER = 2.65\%) with a homograph accuracy of 94\%.  The code\footnote{\url{https://github.com/speechbrain/speechbrain}} and the pretrained model \footnote{\url{https://huggingface.co/speechbrain}} are available on SpeechBrain \cite{SB2021}. 
We also release the new \textit{LibriG2P} dataset that combines data from LibriSpeech Alignments \cite{lugosch2019speech} and the Wikipedia Homograph\cite{homograph-contextual-word-embeddings} on HuggingFace \footnote{\url{https://huggingface.co/datasets/flexthink/librig2p-nostress-space}}.

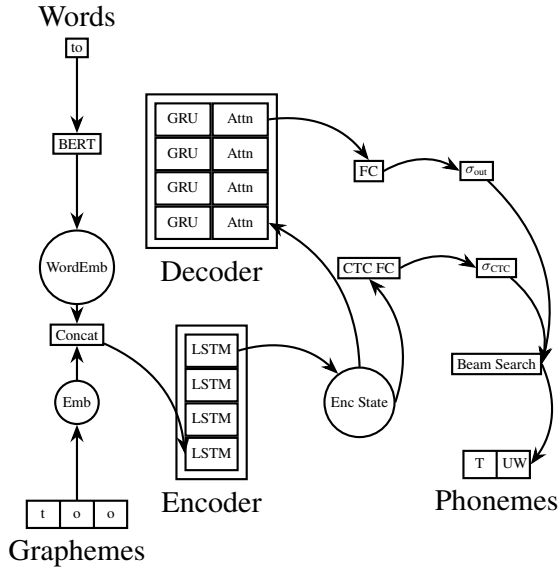
\begin{figure}[t!]
    \centering
    \begin{tikzpicture}
     [
        ->,auto,thick,
        scale=1.25, every node/.style={scale=1.25},
        basic/.style={draw, text centered,font=\small,scale=0.5},
        circ/.style={basic, circle},
        rect/.style={basic, text width=1.5em, text height=1em, text depth=.5em},
        compact/.style={rect, text width=3em},    
        encoder/.style={basic, text width=2.8em, text height=1em, text depth=.5em},
        decoder/.style={basic, text width=6em, text height=1em, text depth=.5em},
        graphemes/.style={basic, text width=1.5em, rectangle split, rectangle split horizontal=true, rectangle split parts=3},
        phonemes/.style={basic, text width=1.5em, rectangle split, rectangle split horizontal=true, rectangle split parts=2},
        >={Stealth[]}
      ]
      \node [label=below:Graphemes,graphemes] (graphemes) {\strut\text{t}\nodepart{two}\strut\text{o}\nodepart{three}\strut\text{o}};
      \node [circ,above=of graphemes] (emb) { Emb };
      \matrix [encoder,label=below:Encoder,right=of emb] (encoder) {
        \node[anchor=center,basic] (enclstm1) {LSTM}; \\
        \node[anchor=center,basic] (enclstm2) {LSTM}; \\
        \node[anchor=center,basic] (enclstm3) {LSTM}; \\
        \node[anchor=center,basic] (enclstm4) {LSTM}; \\
      };
      \node [circ, right=of encoder] (encstate) {\text{Enc State}};
      \matrix[decoder,label=below:Decoder,above=of encoder] (decoder) {
        \node[anchor=center,compact] (decgru1) {\text{GRU}};& \node[anchor=center,compact] (decattn1) {\text{Attn}}; \\   
        \node[anchor=center,compact] (decgru2) {\text{GRU}};& \node[anchor=center,compact] (decattn2) {\text{Attn}}; \\   
        \node[anchor=center,compact] (decgru3) {\text{GRU}};& \node[anchor=center,compact] (decattn3) {\text{Attn}}; \\   
        \node[anchor=center,compact] (decgru4) {\text{GRU}};& \node[anchor=center,compact] (decattn4) {\text{Attn}}; \\   
      };
      \node [circ, above=of emb] (wordemb) {\text{WordEmb}};                  
      \node [basic, above=of wordemb] (bert) {\text{BERT}};      
      \node [basic, label=above:Words, above=of bert] (words) {\text{to}};      
      \node [basic, at=($(emb)!0.5!(wordemb)$)] (concat) {\text{Concat}};            
      \node [basic, right=of decoder] (lin) {\text{FC}};
      \node [basic,below=of lin] (ctc_lin) {\text{CTC FC}};
      \node [basic,right=of lin] (lin_softmax) {$\sigma_{\text{out}}$};
      \node [basic,right=of ctc_lin] (ctc_lin_softmax) {$\sigma_{\text{CTC}}$};
      \node [basic, below=of ctc_lin_softmax] (beam) {\text{Beam Search}};          
      \node [label=below:Phonemes,phonemes,below=of beam] (phonemes) {\strut\text{T}\nodepart{two}\strut\text{UW}};
    
    \path[every node/.style={font=\sffamily\small}]
        (graphemes) edge node [right] {} (emb)
        (words) edge node [right] {} (bert)
        (bert) edge node [right] {} (wordemb) (emb) edge node [right] {} (concat)  
        (wordemb) edge node [right] {} (concat)          
        (concat) edge [bend left] node [left] {} (enclstm4.west)
        (enclstm1.east) edge [bend left] node [left] {} (encstate)
        (encstate.north) edge [bend right] node [right] {} (decattn4.east)
        (decattn1.east) edge [bend left] node [right] {} (lin.north)
        (lin.east) edge [bend left] node [right] {} (lin_softmax.west)
        (ctc_lin.east) edge [bend left] node [right] {} (ctc_lin_softmax.west)
        (encstate.east) edge[bend right] node [right] {} (ctc_lin.south)
        (beam.east) edge [bend left] node [right] {} (phonemes.east)
        (ctc_lin_softmax) edge [bend left] node [right] {} (beam.east)
        (lin_softmax) edge [bend left] node [right] {} (beam.east)        
        ;
    \end{tikzpicture}
    \caption{Encoder-Decoder Architecture of SoundChoice.}
    \label{fig:architecture_rnn}
\end{figure}

\section{Model Architecture}
The basic architecture of SoundChoice is depicted in Fig. \ref{fig:architecture_rnn}. The input graphemes (discrete) are first encoded into continuous vectors using a simple lookup table that stores embeddings of a fixed dictionary and size.  At this stage, we also combine word-level embeddings from a pretrained BERT model \cite{bert}. This addition inflates higher-level semantic information into the system that improves homograph disambiguation.

An LSTM-based encoder then scans the input characters and derives latent representations that embed short and long-term contextual information. 
On top of the encoder, we use CTC loss (after applying a softmax classifier). %
The encoded states feed a GRU decoder coupled with a content-based attention mechanism. Special tokens called $\langle bos\rangle$ and $\langle eos\rangle$ are used to mark the beginning and end of a sentence, respectively.  On top of the decoder, we combine the standard Negative Log-Likelihood (NLL) loss with our homograph loss. Finally, a hybrid beamsearch mechanism that exploits both the CTC and final predictions is employed. The partial hypotheses are rescored with an RNN language model that operates at the phoneme level.

We will provide more details on the proposed architecture in the following sub-sections.

\subsection{Word Embeddings}
To improve homograph disambiguation, we need our model to learn latent representations that correlate with grammar and semantics knowledge.
We thus hypothesize that features from a large language model trained on a large corpus can improve performance. Although many of the recently-proposed language models could fit our purpose, we here used word embeddings derived from the popular BERT model \cite{bert}.
The BERT embeddings pass through a simple encoder consisting of a normalization layer, a  single downsampling linear layer, and $\tanh$ activation. These features are then concatenated with the character-level embeddings to form a single embedding vector.

\subsection{Tokenization}
We use the SpeechBrain \cite{SB2021} implementation of the SentencePiece \cite{sentencepiece} language-independent tokenizer with a unigram model. The goal is to shorten the grapheme and phoneme sequences, making them easier for the neural network to model. The tokenizer achieves this by learning a transformation to a newly constructed vocabulary comprised of the original tokens and common combinations of tokens encountered in the corpus. As we will discuss in Sec.\ref{sec:res}, we find that tokenization in the character and phoneme spaces is not always helpful and does not play an important role as expected.

\subsection{Encoder/Decoder Architecture}
The encoder and decoder use recurrent neural networks. The encoder is based on an LSTM, while the decoder uses a GRU model coupled with content-based attention. The hyperparameters of the model are shown in Table \ref{tab:hyperparams_rnn}. We derived them by performing a hyperparameter search with Oríon \cite{orion}, where we search for the embedding dimensions, depths, and the number of neurons that maximize the PER on the validation set.

We attempted a variation of this model using residual convolutional layers \cite{conv-g2p}. In particular, we replaced the Bi-LSTM model with a series of residual convolutional layers. We achieve similar performance to the baseline one on lexicon data; however, it does not appear to benefit from pretraining, performing poorly on sentence data.
We also conduct experiments to compare the RNN-based architecture with a Transformer-based one \cite{transformer}. In this case, we use a conformer as an encoder and a standard transformer for decoding. The best hyperparameters are shown in Table \ref{tab:hyperparams_xformer}. As we will see in Sec. \ref{sec:res}, this model performs well but is slightly worse than the aforementioned RNN-based architecture.

\begin{table}[t!]
\small
  \caption{RNN Model Hyperparameters.}
  \label{tab:hyperparams_rnn}
  \begin{tabular}{c c c}
    \toprule
    \textbf{Component} & 
    \textbf{Layer} &
    \textbf{Details} \\
    \midrule
          Embedding &  & Dim = 512 \\
          Encoder & LSTM & 4 layers, 512 neurons, dropout = 0.5 \\
          Decoder & GRU & 4 layers, 512 neurons, dropout = 0.5 \\
          FC & Linear & 43 neurons \\
          CTC FC & Linear & 43 neurons \\
    \bottomrule
  \end{tabular}
\end{table}

\begin{table}[t!]
\small
    \caption{Transformer (Conformer) Model Hyperparameters.}
    \label{tab:hyperparams_xformer}
    \centering
    \begin{tabular}{c c c}
         \hline \\
          \textbf{Component} & \textbf{Layer} & \textbf{Details}  \\
          \hline \\
          Embedding &  & Dim = 256 \\
          Encoder & Convolutional & 2 layers, kernel size=15 \\
          Decoder & Transformer & 2 layers, dim = 4096 \\
          FC & Linear & 43 neurons \\
          CTC FC & Linear & 43 neurons \\
         \hline
    \end{tabular}
\end{table}

\subsection{Beam Search and Language Model}
We employ a hybrid beamsearcher similar to those used in modern speech recognizers \cite{watanabe}. It combines the log probabilities derived from the CTC encoder with those estimated by the decoder. The beamsearcher rescores the partial hypothesis with a phoneme language model. We hypothesize that a language model trained on sequences of phonemes can help minimize uncertainty by choosing the most likely phoneme sequences where an accurate prediction is difficult to make.
We use an RNN-based language model with an embedding dimension of 256 and 2 hidden layers of 512 neurons each, regularized via dropout at a rate of 0.15.

\section{Training}
In this section, we provide more information on the adopted training strategy. 

\subsection{CTC Loss}
The CTC loss is computed on top of the encoder. CTC is suitable for grapheme-to-phoneme because the length of the phoneme sequence does not normally exceed the length of the input characters.
This condition holds for many languages.  For instance, in most European languages, a single grapheme can produce one phoneme by itself, be silent or be part of an n-graph. Languages producing more than one phoneme for single grapheme are rare. One exception is Ukrainian, where the letters "$\epsilon$" and "ï" yield two-phoneme combinations  [\textipa{jE}] and [\textipa{ji:}], respectively. In such cases, the limitation can be addressed via sequence padding or by introducing quasi-categories where a single position stands for two phonemes.

The CTC loss is combined with standard NLL loss used on top of the decoder. This multi-task learning approach improves performance and helps the model convergence significantly. 

\subsection{Homograph Loss}
In a typical ambiguous sentence from Wikipedia Homograph Data \cite{google-homograph-repo}, the homograph only represents an insignificant portion of the whole sentence. An error in the homograph can involve only one or two phonemes out of ~30-250 phonemes in the sentence. This incidence is comparable to random variations in labeling or infrequent, ambiguous, or challenging sequences, such as proper names or acronyms. A model can thus achieve a low PER without successfully disambiguating the homographs.

We mitigate this issue by adding a special loss that amplifies the contribution of the homographs relative to other words in the sentence. This is realized by computing the NLL loss on the subsequence corresponding to the homograph only.

The total loss used to train the G2P is thus a combination of three objectives: 
\begin{equation}
\mathcal{L} = \mathcal{L}_{NLL} + \lambda_{h} \textrm{L}_h + \lambda_{c} \textrm{L}_{CTC} 
\end{equation}

where $\mathcal{L}_{NLL}$, $\mathcal{L}_{h}$, and $\textrm{L}_{CTC}$ are the sequence, homograph, and CTC losses, respectively. The factors $\lambda_{h}$ and $\lambda_{c}$ are used to weight the homograph and CTC losses.

\subsection{Curriculum Learning}
We employ a curriculum learning strategy based on different stages of increasing complexity. First, we learn how to convert words into phonemes using the lexicon information. This step is relatively easy as it involves short sequences without the need to disambiguate homographs. 
In our case, we trained the model with this modality for 50 epochs. Then, we move training on by considering whole sentences from LibriSpeech-Alignments \cite{lugosch2019speech}. This step is more challenging, but the model pretrained on single words is already well-initialized for addressing this task. We train the model on sentences for  35 epochs.
Finally,  we perform a fine-tuning step using the homograph dataset for up to 50 epochs. 

The adoption of this curriculum learning strategy turned out to play a crucial role in our G2P system. Without using it, the system provides a worse performance and struggles to converge. Further details on training can be found in the released code\footnote{\url{https://github.com/speechbrain/speechbrain/tree/develop/recipes/LibriSpeech/G2P}}.

\section{Experimental Setup}
\subsection{Datasets}
\label{sec:dataset}
We train the Grapheme-to-Phoneme Model  using LibriSpeech-Alignments\cite{lugosch2019speech},  Google Wikipedia Homograph Data\cite{gorman-etal-2018-improving}\cite{google-homograph-repo} and CMUDICT\cite{cmudict}.

The set of outputs consists of 41 phonemes (ARPABET without stress markers) plus a word-separator token. The original phoneme annotations in LibriSpeech-Alignments \cite{lugosch2019speech} lack a word separator; its position is inferred from the word-level annotation. Google Wikipedia Homograph Data \cite{gorman-etal-2018-improving}\cite{google-homograph-repo}, instead, lacks the phoneme annotations completely. However, each sample is tagged for the type of homograph it includes. Phoneme annotations are constructed by searching the tagged homograph in the provided glossary (with phonemes mapped from IPA to ARPABET) and looking up the remaining words in CMUDICT\cite{cmudict}. Uppercase words appearing in the original text that do not exist in CMUDICT \cite{cmudict} are interpreted as acronyms. We drop samples where the aforementioned methods fail.

We construct a new combined dataset named LibriG2P specialized for G2P with 3 slices: a \textit{lexicon} consisting of each unique word encountered in LibriSpeech\cite{librispeech} as a separate sample, a \textit{sentence} slice consisting of entire LibriSpeech dataset annotated for phonemes derived from Librispeech-Alignments\cite{lugosch2019speech} and a \textit{homograph} slice consisting  of a subset of the Wikipedia Homograph\cite{gorman-etal-2018-improving}\cite{google-homograph-repo} dataset. The non-space-enabled version lacks the homograph slice because the underlying implementation relies on word boundaries to locate the homograph. The train-validation-test split follows Table \ref{tab:data_splits}.
\begin{table}[]
\small
    \centering
    \begin{tabular}{c c c c c}
        \hline
         \textbf{Type} & \textbf{Train} & \textbf{Validation} & \textbf{Test} & \textbf{Total} \\
         \hline
         Lexicon & 202377 & 2065 & 2066 & 206508 \\
         Sentence & 103967 & 2702 & 2702 & 109371 \\
         Homograph & 9231 & 516 & 512 & 10259 \\
         \hline
    \end{tabular}
    \caption{LibriG2P splits.}
    \label{tab:data_splits}
\end{table}
The Google Wikipedia Homograph\cite{gorman-etal-2018-improving}\cite{google-homograph-repo} dataset is highly unbalanced with regards to the frequencies of pronunciation variations for any given homograph. We conduct experiments both with random sampling ("unbalanced") and with weighted sampling attempting to equalize the probability of each variation being selected ("balanced").

Given the inconsistency between LibriSpeech-Alignments\cite{librispeech} annotations obtained from audio and annotations computed using CMUDict\cite{cmudict}, primarily in unstressed syllables or short connecting words - conjunctions and prepositions (e.g. "and": [\textipa{@nd}] vs [\textipa{\ae nd}] or into - [\textipa{intu:}] vs [\textipa{int@}], we produce a variation of the dataset with non-homograph words in the \textit{sentence} slice replaced with CMUDict\cite{cmudict} pronunciations where possible.

To foster replicability and follow-up studies, we release the processed datasets to the community.

\subsection{Metrics}
We use the Phoneme Error Rate (PER\%) to evaluate all models. To evaluate the performance of homograph disambiguation, we compute an additional metric, the \textit{homograph classification accuracy}, defined as the percentage of samples in which the pronunciation of the homograph is predicted with no errors. 

\section{Results}
\label{sec:res}

\subsection{RNN Model}
Table \ref{tab:model_results_rnn} reports the performance achieved with the RNN model under different settings. 
It clearly emerges that sentence-based systems significantly outperforms word-based systems (see row 1 vs row 2 of Table \ref{tab:model_results_rnn}). This change leads to a relative improvement of 47\% in the PER,  confirming the key importance of contexts in grapheme-to-phoneme conversion. Table  \ref{tab:model_results_rnn} also highlights the importance of adding a special token (space) in the phoneme space. This token is needed to signal word boundaries and injects prior information about words into the system. This simple trick leads to a further 18\% relative improvement of the PER. The tokenizer applied to phonemes leads to a minor performance improvement, while the BERT word embeddings do not improve the PER. BERT embeddings, however, will play a crucial role in homograph disambiguation. The phoneme language model turned out to not play a significant role as well. The best system achieves a PER of 2.65\% on the LibriSpeech dataset.

\subsection{Transformer Model}
Table \ref{tab:model_results_xformer} reports the results achieved with the Conformer/Transformer model. The important benefits observed using a sentence-based system and adding the space token are confirmed. The minor role played by BERT embeddings and language models is observed for this model as well. In terms of performance, the best RNN model outperforms the transformer one (PER=2.65\% vs PER=2.83\% ). The performance drop is not huge and might be because transformers 
notoriously require large datasets to be trained properly.

\subsection{Homograph Disambiguation}
Table \ref{tab:results_hgraph} reports the results achieved after fine-tuning the model with the homograph dataset.  The homograph disambiguation is evaluated on variations of the best-performing RNN model, with and without word embeddings. The use of the proposed homograph loss improves the homograph accuracy. With a weight factor $\lambda_h$ of 2.0, the accuracy improves from 82\% (no homograph loss) to 87\%, thus corroborating our conjecture that the signal from the NLL loss on the entire sentence alone is not strong enough for disambiguation.
BERT\cite{bert} embeddings significantly improve homograph detection as well. Thanks to this addition, the best system reaches an accuracy of 94\% in homograph disambiguation. While the disambiguation accuracy cited in \cite{gorman-etal-2018-improving} is higher, this method achieves competitive results within the sequence model itself without additional classifiers.

It is worth mentioning that reevaluation of LibriSpeech\cite{librispeech} data after homograph fine-tuning showed a deterioration in nominal PER. This is due to inconsistencies in labeling, given that LibriSpeech Alignments was annotated using an automated aligner, capturing minor subtleties in pronunciation, whereas the homograph step relied on a dictionary\cite{cmudict} and allowed for only one pronunciation per word except for the homograph. When reevaluating on LibriSpeech Alignments after homograph fine-tuning, the test PER increases from 2.65\% to 4.20\%. Qualitative analysis reveals that most of the new errors originate from allowable variations in the labeling of non-homograph words, especially in prepositions/conjunctions and unstressed vowels. Retraining the model on the version of the dataset where the original labels are harmonized with CMUDict\cite{cmudict} leads to an overall PER decrease to 1.54\%. This suggests that the apparent error increase in fine-tuning is due to a distribution shift rather than catastrophic forgetting.

 \begin{table}[t!]
 \small
    \centering
    \setlength\tabcolsep{1.8pt}
    \begin{tabularx}{200pt}{c c c c c c c c c}
        \hline
         \# & \textbf{Sentence}  & \textbf {Space} & \textbf{TP} &   \textbf{WE} & \textbf{LM} & \textbf{Val PER} & \textbf{Test PER}  \\
         \hline
         1 & &  &  &  &  & 6.82 & 6.46 \\
         2 & \checkmark &  &  &  & & 3.23  & 3.38     \\
         3 & \checkmark & \checkmark  &  &  & & 2.63 & 2.76 \\
         4 & \checkmark & \checkmark  & \checkmark  &  &  & 2.56 & 2.69 \\
         5 & \checkmark & \checkmark & \checkmark  & \checkmark &  & \textbf{2.42} & 2.71 \\
         6 & \checkmark & \checkmark &  &  & \checkmark & 2.51 & \textbf{2.65} \\
         \hline
    \end{tabularx}
    \caption{G2P Model Results - RNN. \textbf{Sentence} is flagged when training/evaluating on full sentences, \textbf{Space} refers to the space token preserved. \textbf{TP} is marked when applying the tokenization to phonemes, \textbf{WE} refers to BERT embeddings, while \textbf{LM} is flagged when the phoneme language model is used.}
    \label{tab:model_results_rnn}
\end{table}
    
\begin{table}[t!]
\small
    \centering
    \setlength\tabcolsep{1.8pt}
    \begin{tabularx}{200pt}{c c c c c c c c}
         \hline
         \# & \textbf{Sentence}  & \textbf {Space} & \textbf{TP} &   \textbf{WE} & \textbf{Val PER} & \textbf{Test PER}  \\
         \hline
         1 & &  &  &  &  9.11  & 9.23 \\
         2 & \checkmark &  &   &  & 5.30 & 5.46 \\
         3 & \checkmark & \checkmark &  &  &  3.59 & 3.70 \\
         4 & \checkmark & \checkmark & \checkmark  &  &  \textbf{2.74} & \textbf{2.83} \\
         5 &\checkmark & \checkmark & \checkmark & \checkmark & 2.79 & 2.97 \\  
         \hline
    \end{tabularx}
    \caption{G2P Model Results - Conformer. \textbf{Sentence} is flagged when training/evaluating on full sentences, \textbf{Space} refers to the space token preserved. \textbf{TP} is marked when applying the tokenization to phonemes, \textbf{WE} refers to BERT embeddings, while \textbf{LM} is flagged when the phoneme language model is used.}
    \label{tab:model_results_xformer}
\end{table}

\begin{table}[t!]
\small
    \centering
    \begin{tabularx}{200pt}{c c c c c }
         \hline
         & \textbf{Word Emb} & \textbf{HG Weight} &  \textbf{Bal} & \textbf{Accuracy} \\
         \hline
         1 &  & 0.0 &  & 82 \% \\
         2 &  & 2.0 &  & 87 \% \\
         3 &  & 2.0 & \checkmark & 85 \% \\
         4 &  & 5.0 & \checkmark & 82 \% \\
         5 & \checkmark & 2.0 & \checkmark & \textbf{94} \% \\
         \hline
    \end{tabularx}
    \caption{Homograph Disambiguation Results. \textbf{Word Emb} refers to Word embeddings, \textbf{HG Weight} is the weight of the homograph loss ($\lambda_h$), while \textbf{Bal} refers to balanced sampling.}
    \label{tab:results_hgraph}
\end{table}
\FloatBarrier

\section{Conclusions}
This work proposed SoundChoice, a novel method for converting grapheme-to-phonemes that is robust against homograph disambiguation.  The model is trained with a curriculum learning strategy that learns a word-based system first and finally learns a sentence-based model with a special homograph disambiguation loss. The best solution relies on an RNN system with hybrid/CTC attention and beam search. 
It takes advantage of word embeddings from a pre-trained BERT model as well. We achieved a PER of 2.65\% on whole-sentence transcription using data from LibriSpeech and 94\% accuracy in homograph detection using the Google Wikipedia Homograph\cite{gorman-etal-2018-improving}\cite{google-homograph-repo} corpus.

SoundChoice can be used in different ways in speech processing pipelines. For instance, it allows the training of TTS systems with phoneme tokens (or with mixed representation \cite{kyle1}). It can be used for speech recognition as well, as phonemes are known to be excellent targets, especially in channeling scenarios where speech is corrupted by noise and reverberation \cite{robust1, robust2}. 
In future work, we would like to extend this approach to address multiple languages.

\bibliographystyle{IEEEtran}

\bibliography{mybib}

\end{document}